\documentstyle[prl,aps,twocolumn,psfig]{revtex}
\begin{document}
\title{Implementation of a 
Deutsch-like quantum algorithm utilizing
entanglement at the two-qubit level, on an
NMR quantum information processor}
\author{Kavita Dorai$^1$~\cite{kav-email},
Arvind$^3$~\cite{arv-email},
and Anil Kumar$^{1,2}$~\cite{ak-email}}
\address{$^1$Department of Physics, and 
$^2$Sophisticated Instruments Facility, 
Indian Institute of Science, Bangalore 560012 India}
\address{$^3$Department of Physics, 
Guru Nanak Dev University, Amritsar 143005 India}
\maketitle 
\begin{abstract}
We describe the experimental implementation of
a recently proposed quantum algorithm involving
quantum entanglement at the level of two qubits using
NMR. The algorithm solves a generalisation of the
Deutsch problem and distinguishes between even and
odd functions using fewer function calls than is
possible classically. The manipulation of entangled
states of the two qubits is essential here, unlike
the Deutsch-Jozsa algorithm  and the Grover's search
algorithm for two bits.  
\end{abstract}
\section{Introduction}
It has been demonstrated recently that a quantum
computer, exploiting quantum state superposition
and entanglement, is definitely more powerful
than any existing classical 
computer~\cite{divin-95}.
Many ``quantum'' algorithms that achieve the
same computational tasks much faster than their
classical counterparts have been 
designed~\cite{deu-roy-92,cleve-98,shor-97,grover-97}.
Thus far, NMR has been the most successful method employed
to physically implement small quantum information
processors and to test the power of intrinsically
quantum 
algorithms~\cite{gersh-sci-97,cory-proc}. 
The visualisation of a spin-1/2 particle as a qubit,
combined with existing multi-dimensional NMR
methods has led to  breakthroughs in 
pseudopure state 
preparation~\cite{cory-proc,knill-pra-98,van,kav-pra}, 
the demonstration of universal quantum logic
gates~\cite{cory-physica,nielsen,j-jmr,mad,j-jmr2}, 
and the 
implementation of various quantum 
algorithms~\cite{j-jcp,ch-nat,ch-prl,j-nat,j-sci,lin,marx,j-prl,tsm-2d}.

Early on, Deutsch and Jozsa presented a simple
problem to determine whether a Boolean
function $f\/$
is constant or balanced~\cite{deu-roy-92}.
Classically, the algorithm requires many function
calls to solve the problem without error, but a
quantum computer can solve the problem using only
a single function call. The Cleve version of the
algorithm used an extra qubit to encode the
required unitary transformations and the solution
could be read out as the relative phase of the
qubits~\cite{cleve-98}. Recently, the
Deutsch-Jozsa (DJ) problem was modified and
implemented using a lesser number of 
input qubits~\cite{collins-pra-98,arv-dj,korean-dj}, and interest 
continues in further generalisations of 
the DJ problem~\cite{korean-theory}. It was realised
that, for upto two bits the DJ problem need not
invoke entangling transformations for its
solution. The problem thus allows a classical
description for the two-qubit case, and it is
only for three or more qubits that the quantum
nature of the algorithm is displayed~\cite{collins-pra-98,arv-dj}.
As pointed out recently by Seth Lloyd~\cite{lloyd-grover}, 
Grover's search algorithm for two qubits also does not require
entangling transformations. Hence both these algorithms become
truly quantum only for three or more qubits.

A quantum algorithm to distinguish between even and odd
functions using fewer function calls than
a classical algorithm has been 
recently designed~\cite{arv-qph-00}. The
algorithm uses entangling transformations at the
two-qubit level itself and is an interesting
example of the power of a quantum computer over
corresponding classical systems. 
In this paper,
we present the experimental implementation of
the algorithm on a two-qubit NMR quantum
computer. The requisite unitary transformations
have been implemented using spin-selective and
composite-z pulses. A judicious combination of
composite pulses and evolution under the scalar
coupling Hamiltonian has been used to construct
the desired entangling transformations. 
The algorithm requires distinguishing between non-orthogonal
states of the two qubits in order to classify the 
functions~\cite{non-ortho}.
In our NMR implementation, such a distinction is
achieved in a single measurement.
The algorithm to evaluate the even or odd nature of
a function uses entangling transformations for its
implementation on two input qubits.
Entangling transformations can produce entangled
states that have no classical 
analogue~\cite{ekert-amj-94}.

Consider a Boolean function defined from a two-bit domain 
space to a one-bit
range space: $f(x): \{ 0,1 \}^{2} \rightarrow \{ 0,1 \}\/$.
There are four possible input values $(00),(01),(10)$
and $(11)$, the output for each of these being either 
$0\/$ or $1\/$. The 16 possible functions can be
divided into sub-classes based on the number of
ones and zeros in their outputs. 
The functions can be
categorised in the sub-classes
$[0,4], \, [1,3], \,[2,2], \,[3,1], $ or \,$
[4,0]\/$, where the first entry 
indicates the number of ones and the second indicates the number of 
zeros in the output. The functions with an even number  of 
ones in the output (the functions belonging to
the categories  $[0,4], \, [2,2]\/$ and $[4,0]\/$)  are denoted 
``{\bf even}'' functions while the functions with 
an odd number of ones in the output (the $[1,3]$ and $[3,1]$ functions) 
are said to be  ``{\bf odd}''
functions. For the two-qubit case, we thus have 8
even and 8 odd functions. 

Classically, the classification of a function 
as even or odd would require 
computing it at all input points. The quantum
algorithm constructed uses just two function
calls to evaluate the even or odd character of the
given function~\cite{arv-qph-00}. The implementation requires a
quantum gate to call the function, and a judicious use of 
pseudo-Hadamard transformations (on both qubits, and
selectively on one qubit alone).
The function call mechanism is similar to the one
used to solve the modified 
Deutsch problem~\cite{collins-pra-98,arv-dj,korean-dj}. 
Each function $f\/$ can be encoded by a unitary
transformation $U_f\/$, 
with its action on the eigenstates of the two qubits
being defined as
\begin{eqnarray}
&\vert x \rangle_{\mbox{\tiny 2-bit}}
\stackrel{U_{f}}{\longrightarrow}
(-1)^{f(x)} \vert x \rangle_{\mbox{\tiny 2-bit}}&
\nonumber \\
\nonumber \\
&
U_f=\left(
\begin{array}{cccc}
(-1)^{f(00)}&0&0&0\\
0&(-1)^{f(01)}&0&0\\
0&0&(-1)^{f(10)}&0\\
0&0&0&(-1)^{f(11)}
\end{array}
 \right)
&
\end{eqnarray}
Pseudo-Hadamard
gates~\cite{j-jmr} are practically equivalent
to the Hadamard operator 
and have been utilised throughout our analysis. The one-qubit 
pseudo-Hadamard gate is given by
\begin{eqnarray}
&
\begin{array}{c}
\vert 0\rangle \stackrel{h}{\rightarrow} 
\frac{\scriptstyle 1}{\scriptstyle \sqrt{2}}
(\vert 1 \rangle + \vert 0 \rangle) \\
\vert 1\rangle \stackrel{h}{\rightarrow} 
\frac{\scriptstyle 1}{\scriptstyle \sqrt{2}}
(\vert 1 \rangle - \vert 0 \rangle)
\end{array}
;\,
h = \frac{1}{\sqrt{2}}
\left(\begin{array}{lr} 
{ 1} & {1}\\
{ -1} & {1}
\end{array}
\right)  &
\end{eqnarray}

In our analysis, the pseudo-Hadamard gate is applied
on both qubits non-selectively as well as
selectively on one qubit alone. The corresponding
gates are denoted as
\begin{equation}
 h^{(1)}=h \otimes I;\, 
 h^{(2)}=I \otimes h;\, 
h^{(1,2)}= h \otimes h 
\label{hadamard}
\end{equation}
where $1\/$ and $2\/$ label the qubit involved.

There are sixteen $U_f\/$ matrices in all, with
eight of them being entangling and the rest
non-entangling in character. 
For example, the $U_f\/$ matrix with diagonal entries
$[1,1,1,-1]\/$ cannot be written as a tensor product of two 
matrices, one belonging to each qubit. This
unitary transformation is hence entangling in nature.
It is interesting to note that for the two-bit case, 
the sub-class of functions that are either constant or balanced in
the sense of the Deutsch problem i.e. the 
functions belonging to the $(0,4)\/$ and
$(2,2)\/$ sub-classes, are all separable in character. 
Therefore, the two-bit Deutsch
problem affords a classical explanation and 
can be implemented using non-entangling transformations alone.

\def\ufbox{\multiput(0,0)(7,0){2}{\line(0,1){20}}
\multiput(0,0)(0,20){2}{\line(1,0){7}} 
\boldmath
\put(1.2,10){$U_f$}
\unboldmath
}
\def\hfbox{\multiput(0,0)(7,0){2}{\line(0,1){7}}
\multiput(0,0)(0,7){2}{\line(1,0){7}} 
\boldmath
\put(2.1,2.5){h}
\unboldmath
}
\def\hfibox{\multiput(0,0)(7,0){2}{\line(0,1){7}}
\multiput(0,0)(0,7){2}{\line(1,0){7}} 
\boldmath
\put(1.1,2.5){h$^{\mbox{-}1}$}
\unboldmath
}
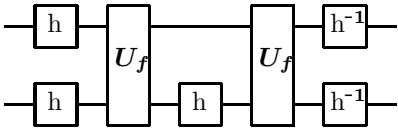
\begin{figure}
\unitlength=.8mm
\begin{picture}(80,30)
\boldmath
\thicklines
\put(15,5){\hfbox}
\put(15,18){\hfbox}
\put(27,5){\ufbox}
\put(39,5){\hfbox}
\put(51,5){\ufbox}
\put(63,18){\hfibox}
\put(63,5){\hfibox}
\multiput(0,8.5)(0,13){2}{
\put(10,0){\line(1,0){5}}
\put(22,0){\line(1,0){5}}
\put(58,0){\line(1,0){5}}
\put(70,0){\line(1,0){5}}
}
\put(34,21.5){\line(1,0){17}}
\put(34,8.5){\line(1,0){5}}
\put(46,8.5){\line(1,0){5}}
\end{picture}
\caption{Quantum circuit for two qubits, that implements the
algorithm to distinguish between even and odd functions,
using pseudo-Hadamard
gates (denoted by $h\/$) and the $U_f\/$ transformations.
Pseudo-Hadamard gates $h\/$ convert an eigenstate to a uniform
superposition of all possible eigenstates while $h^{-1}\/$
implement the inverse operation.}
\label{circuit}
\end{figure}
A quantum circuit to evaluate the even or
odd nature of a function is shown in
Figure~\ref{circuit}. The algorithm starts with
both qubits in a thermal (mixed) initial state.
A pseudo-Hadamard transformation is
applied on both qubits, resulting in both of
them being in a state which is a superposition
of all possible input states.

The desired unitary transformation $U_f\/$ is
then applied on this superposition state,
followed by a pseudo-Hadamard gate operating on
the second qubit alone. The function is called
again through $U_f\/$ and an inverse
pseudo-Hadamard is applied on both qubits before
the read-out operation. The result of the
computation is encoded in the final state of
the two qubits. If the function is even,  both the
qubits are in an unentangled state and a signal
is obtained for the qubit on which the selective
pseudo-Hadamard gate is applied, with no signal
being obtained for the other qubit.
If the function is odd, both
qubits are in an entangled state, corresponding
to multiple-quantum coherence of the two spins, and
no observable signal is obtained.
The even or odd nature of the function can
thus be distinguished ``pictorially'', by
looking at the final NMR spectrum.

Previously, the algorithm has been described for a pure
initial state~\cite{arv-qph-00}. In NMR one
normally encounters mixed states.
Hence we first generalise the procedure to
include mixed initial states. 
Consider the two qubits in an initial state described by the 
deviation density matrix 
\begin{equation}
\Delta \rho_{\rm initial}=
\left(
\begin{array}{cccc}
k_1&0&0&0\\
0&0&0&0\\
0&0&0&0\\
0&0&0&k_2
\end{array}\right)
\end{equation}
with $k_1$ and $k_2$ being independent variables.
For $k_1=1\/$ and $k_2=0\/$ this density matrix represents a 
pseudopure state. For $k_1=1\/$ and 
$k_2=-1\/$ it represents a thermal state. Multiples of identity 
have been ignored in the density matrix, as they do not contribute to
the measured signal in an NMR experiment.

We now evolve this density matrix through the
sequence of operations 
$h^{(1,2)}\, U_f\, h^{(2)}\, U_f\, \left[h^{(1,2)}\right]^{-1}\/$.
After some algebra, this leads to the deviation density matrix 
\begin{eqnarray}
\Delta \rho_{\rm even} = \frac{1}{2}\left(
\begin{array}{cccc}
k_2&\zeta k_2&0&0\\
\zeta k_2&k_2&0&0\\
0&0&k_1&\zeta k_1\\
0&0&\zeta k_1&k_1
\end{array}
\right), \,\, \mbox{and} \nonumber \\
\Delta \rho_{\rm odd} = \frac{1}{2}\left(
\begin{array}{cccc}
k_1+k_2&0&0&\zeta(k_1-k_2)\\
0&0&0&0\\
0&0&0&0\\
\zeta(k_1-k_2)&0&0&k_1+k_2\\
\end{array}
\right)
\end{eqnarray}
where $\zeta = +1\/$ or $-1\/$, for different $U_f\/$ transformations.
For $k_1=1\/$ and $k_2=0\/$ we arrive at the
the pure state result 
of~\cite{arv-qph-00} i.e., the final state
is $\vert 0 0 \rangle + \zeta \vert 0 1 \rangle\/$ for
an {\bf even} function and $\vert 0 0 \rangle + \zeta \vert 1 1 \rangle\/$
for an {\bf odd} function.
For a thermal initial state i.e. 
$k_1=-k_2=1\/$, the deviation density matrices
$\Delta \rho_{\rm even}\/$ and $\Delta \rho_{\rm odd}\/$,
can be distinguished unambiguously by
a single NMR measurement. 
The spectrum for even functions will give
two lines corresponding to the 
observable single quantum coherences present in the density 
matrix.  For odd functions there is no signal as the
only non-diagonal elements of the density matrix
are the ones corresponding 
to double-quantum coherences.
We note here that a single NMR measurement is able
to distinguish between two non-orthogonal quantum
states which is normally not possible using other 
measurement techniques~\cite{non-ortho}.
We further note that the presence of double-quantum
coherences in the final state for the odd
functions shows the entangling nature of the
unitary transformations used.
The use of a Hadamard instead of a pseudo-Hadamard transformation 
would lead to results which are qualitatively similar. The
entanglement in the final state would show up as a zero-quantum
coherence of the two qubits, instead of a double-quantum
coherence~\cite{arv-qph-00}.
We have implemented the two-qubit even/odd 
quantum algorithm using
the molecule of 5-Fluorouracil (dissolved in
DMSO) as an NMR quantum computer, with the
fluorine and the ortho-proton being identified as
the two input qubits. This fluorine-proton spin
system is a good candidate for quantum computing
since it has good sensitivity, a resolved
J-coupling of 6.1 Hz and the duration of spin-selective
pulses can be relatively short. All
experiments have been performed on a Bruker
AMX-400 spectrometer at room temperature.
The pseudo-hadamard gate has been achieved
by applying a $(90^{0})_{y}\/$ pulse
selectively on a spin, or non-selectively
on both spins, as the case maybe.

\def\pbx{\multiput(0,0)(6,0){2}{\line(0,1){10}}
\multiput(0,0)(0,10){2}{\line(1,0){6}} 
\put(1.5,4){$\frac{\displaystyle \pi}{\displaystyle 2}$}
\put(2,11){x}
}
\def\pby{\multiput(0,0)(6,0){2}{\line(0,1){10}}
\multiput(0,0)(0,10){2}{\line(1,0){6}} 
\put(1.5,4){$\frac{\displaystyle \pi}{\displaystyle 2}$}
\put(2,11){y}
}
\def\pbmx{\multiput(0,0)(6,0){2}{\line(0,1){10}}
\multiput(0,0)(0,10){2}{\line(1,0){6}} 
\put(1.5,4){$\frac{\displaystyle \pi}{\displaystyle 2}$}
\put(1,11){-x}
}
\def\ppb{\multiput(0,0)(6,0){2}{\line(0,1){10}}
\multiput(0,0)(0,10){2}{\line(1,0){6}}
\put(1.5,4){$\pi$}
\put(2,11){y}
}
\begin{figure}
\unitlength=0.8mm
\begin{picture}(80,72)
\boldmath
\thicklines
\put(0,70){Even Function U$_4$:}
\multiput(15,40)(0,15){2}{
\put(0,0){\line(1,0){30}}
\put(5,0){\pbx}
\put(17,0){\pbmx}
\put(11,0){\ppb}
}
\put(0,30){Odd Function U$_9$:}
\multiput(15,0)(0,15){2}{
\put(0,0){\line(1,0){47}}
\put(5,0){\pbx}
\put(11,0){\pby}
\put(17,0){\pbmx}
\put(24,3){$\tau/2$}
\put(32,0){\ppb}
\put(40,3){$\tau/2$}
\put(47,0){\line(0,1){2}}
}
\multiput(2,3)(0,40){2}{qubit 2}
\multiput(2,17)(0,40){2}{qubit 1}
\end{picture}
\caption{NMR pulse sequences to implement the 
even function $U_4\/$ and the odd function $U_9\/$.
The pulses are represented by boxes, with the phase of each
pulse written above it. Composite-$z\/$ pulses are implemented
by three pulses, applied back-to-back.
The time period $\tau\/$ is set to $1/2J\/$, $J\/$ being
the value of the spin-spin coupling.
The pulse schemes for the other functions are similar and
can be constructed by varying the phase of the composite-$z\/$ pulses.}
\label{schemes}
\end{figure}
We have used pulse schemes consisting of 
sandwiches of composite and spin-selective
pulses (described schematically in
Figure~\ref{schemes}), to implement the even
and odd $U_f\/$ transformations.

\begin{figure}
\hspace*{1.25cm}
\psfig{figure=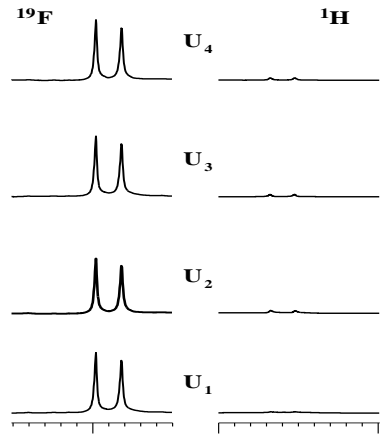,height=6cm,width=5cm,angle=0}
\caption{ The implementation of the even transformations $U_1 - U_4\/$ 
on the two qubits (${}^{19}\/$F and ${}^{1}\/$H) of 5-Fluorouracil.
The  corresponding unitary matrices have entries along the
diagonal of [1,1,1,1], [1,1,-1,-1], [1,-1,1,-1] and
[1,-1,-1,1] respectively. The transformations $U_5 - U_8\/$
($U_{i+4} = - U_{i}\/$, $i=1,2,3,4\/$), lead to exactly
the same results (spectra not shown). All spectra have
been plotted to the same scale.}
\label{evenfig}
\end{figure}

\begin{figure}
\hspace*{1.5cm}
\psfig{figure=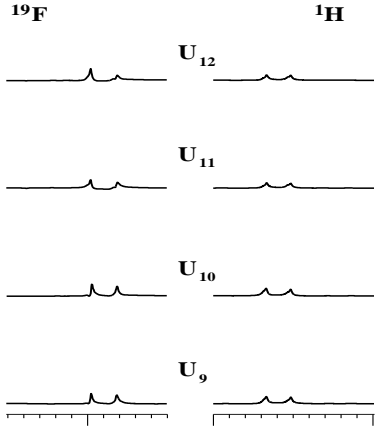,height=6cm,width=5cm,angle=0}
\caption{The implementation of odd functions $U_9 - U_12\/$ on 
5-Fluorouracil, using entangling transformations. The corresponding
unitary matrices have entries along the diagonal of
[1,-1,-1,-1],[-1,1,-1,-1],[-1,-1,1,-1] and [-1,-1,-1,1] respectively.
The odd functions $U_13 - U_16\/$ ($U_{i+4} = U_{i}\/$, $i=9,10,11,12\/$)
lead to the same results (spectra not shown).
A $\tau\/$ period of 82.24 ms has been used to implement the
evolution under the scalar coupling Hamiltonian. All the spectra shown have
been plotted to the same scale as in Fig.~\ref{evenfig}.}
\label{oddfig}
\end{figure}

As an
illustration, the unitary transformation
$U_4\/$ (given by the diagonal matrix 
$[1,-1,-1,1]\/$) corresponds to a
$\pi\/$-rotation about the $z$-axis of both the
qubits, upto a global
phase factor. Global phase changes are not
detectable in NMR and are hence ignored for
the purposes of the experiment. 
The $z$-rotation can be
implemented using a composite-pulse sandwich,
as a set of rotations about the $x\/$ and $y\/$
axes
$[\theta]_z \equiv [\pi/2]_x [\theta]_y [\pi/2]_{-x}\/$
The $U_f\/$'s encoding the other
even functions $U_1\/$, $U_2\/$ and $U_3\/$ correspond to the do-nothing
operation (the transformation is the unity
matrix), a $[\pi]_z\/$ rotation in the
single-spin subspace of the first qubit, and
a $[\pi]_z\/$ rotation on the second qubit respectively, and
have been constructed using similar pulse schemes. 
All these even transformations can be decomposed
into transformations in the subspaces of
each individual qubit and have hence been
implemented experimentally without invoking
quantum entanglement. The result of applying all
the even transformations on the two qubits
is shown in Fig~\ref{evenfig}. Spin-selective
pulses of $12.7 \, \mu\/$ secs on the proton
and $22.1 \, \mu\/$ secs on the fluorine have
been used to achieve good selectivity. The
spectra obtained reveal a retention of the fluorine
spin single-quantum coherence, while no lines
are seen for the proton (whose coherence has
been converted back to unobservable
$z$-magnetization by the selective
pseudo-Hadamard gate). 
We have used entangling
transformations to implement the odd
functions. Consider the non-separable $U_9\/$ matrix with
the entries $[1,-1,-1,-1]\/$ along its diagonal.
The transformation is achieved experimentally
by a $[\pi/2]_z\/$ rotation on the first spin, 
followed by a $[\pi/2]_z\/$ rotation on the second spin,
and then a free evolution for a time interval $\tau\/$
tailored to $\tau= 1/2J\/$, $J\/$ being the
value of the scalar coupling.
A non-selective $\pi\/$ pulse has been applied in the
middle of the $\tau\/$ interval, to refocus the
chemical shift evolution.
The $[\pi/2]_z\/$ rotations in the single-spin
subspaces have been achieved by the composite
pulse sandwiches $[\pi/2]_x [\pi/2]_y
[\pi/2]_{-x}\/$, selective on the spin concerned.
The other odd functions have been similarly
implemented. The spectra corresponding to the
experimental implementation of the odd functions
is shown in Fig.~\ref{oddfig}. The states of
the two qubits are completely entangled, leading
to no observable spectral lines for all the
these functions. All the spectra in both Figs.~\ref{evenfig}
and~\ref{oddfig} have been plotted to the same scale.   

In conclusion, we have shown the NMR
implementation of a quantum algorithm
that uses the entangled states
of two input qubits to distinguish between
even and odd functions. We have used
spin-selective, composite pulse sandwiches
and evolution under scalar coupling to
implement the required non-trivial entangling
transformations. It is interesting 
that NMR experiments are able to
distinguish between two non-orthogonal 
quantum mechanical states in a single measurement.

SIF at
IISc Bangalore
is acknowledged for the use of the AMX-400 spectrometer and IISc for
financial support.

\end{document}